\documentclass[10pt,letterpaper]{article}

\usepackage{cogsci}
\usepackage{pslatex}
\usepackage{apacite}
\usepackage{amsmath}
\usepackage{amsfonts}
\usepackage{graphicx}

\title{Semantic Compression of Episodic Memories}

\author{{\large \bf David G. Nagy\textsuperscript{1,2}, Bal\'{a}zs T\"{o}r\"{o}k \textsuperscript{1,3}, Gerg\H{o} Orb\'{a}n \textsuperscript{1}} \\
  \{nagy.g.david, torok.balazs, orban.gergo\}@wigner.mta.hu \\
  \textsuperscript{1}Computational Systems Neuroscience Lab, MTA Wigner Research Centre for Physics, Budapest, Hungary \\
  \textsuperscript{2}Institute of Physics, E\"{o}tv\"{o}s Lor\'{a}nd University, Budapest, Hungary \\
  \textsuperscript{3}Department of Cognitive Science, Budapest University of Technology and Economics, Budapest, Hungary}

\begin{document}

\maketitle

\begin{abstract}
Storing knowledge of an agent's environment in the form of a probabilistic generative model has been established as a crucial ingredient in a multitude of cognitive tasks. Perception has been formalised as probabilistic inference over the state of latent variables, whereas in decision making the model of the environment is used to predict likely consequences of actions. Such generative models have earlier been proposed to underlie semantic memory but it remained unclear if this model also underlies the efficient storage of experiences in episodic memory. We formalise the compression of episodes in the normative framework of information theory and argue that semantic memory provides the distortion function for compression of experiences. Recent advances and insights from machine learning allow us to approximate semantic compression in naturalistic domains and contrast the resulting deviations in compressed episodes with memory errors observed in the experimental literature on human memory. 

\end{abstract}

\section{Introduction}

Given the physical constraints on memory resources for the human brain, verbatim storage of all sensory experience is unfeasible. The normative framework for analysing this problem is provided by information theory, where efficient compression into memory traces hinges on the agent being able to prioritise information according to its relevance. In information theory these priorities are represented by the distortion function, which characterises the degree to which a particular form of distortion of the original experience is acceptable to the agent. While the distortion function is a critical component of efficient compression, the theory is agnostic about its specific form. Thus, efficient compression raises the question: What is the appropriate distortion function for human memory, that is, from a continuous stream of experience, how does the human brain determine what to remember and what to forget?

Storing knowledge of an agent's environment in the form of a probabilistic generative model has been established as a crucial ingredient in a multitude of cognitive tasks. Perception can be understood as probabilistic inference over the states of latent variables, where prior knowledge is integrated with noisy and ambiguous observations \cite{Kersten2004ObjectInference}. In decision making, the model of the environment is used to predict likely consequences of actions, enabling the agent to find the action policy leading to maximal rewards \cite{Sutton1998a}. Learning is readily formalised as building a probabilistic model of the environment based on observations. Following previous research, we consider establishing a statistical model of the environment to be the domain of semantic memory and formalise it as a probabilistic generative latent variable model of the environment \cite{Kali2004Off-lineInteractions,Hemmer2009,Nagy2016}. 

In addition to maintaining a probabilistic model in semantic memory, previous research has pointed out that retaining rich representations of specific experiences is also necessary \cite{Nagy2016,Kumaran2016,Lengyel2009}. This form of memory, usually termed episodic memory, is an expensive representational format, which requires compression. Here, we are concerned with the nature of this compression. We argue that in the case of memory, similar to perception, it is the inferences regarding the causes underlying sensory experience that is most relevant for the organism, which is precisely the information captured by the latent variables of semantic memory. Therefore, the information contained in the latent variables is what memory should prioritise when resources are constrained. In formal terms, we propose that semantic memory underlies the compression of episodes through providing the distortion function for episodic memory, a process we term semantic compression.

Empirically, the distortion function of an information compressing system becomes apparent in the pattern of memory errors that it produces. In the case of human memory, an extensive body of work has shown that it is indeed far from a carbon copy of sensory experience. Rather than being random noise however, memory errors show robust and systematic biases. Such systematic biases are thought to reflect rational adaptations to computational resource constraints \cite{Schacter2011}. Making this assumption explicit, we formalise semantic compression in the normative framework of lossy compression. This formalisation provides an opportunity for a unifying normative explanation of a wide variety of memory effects. Recent advances in machine learning yielded efficient tools to learn generative models of complex stimuli. In this study, we harness these advances to compare biases of humans in a recall task using naturalistic sketch images with distortions introduced by semantic compression.

In the following, we give a short overview of the theoretical framework we use in formalising semantic compression: we introduce the information theoretical framework for lossy compression called rate distortion theory, define semantic compression formally, and show how it can be approximated through variational inference, highlighting a recently developed correspondence with approximate methods in rate distortion theory. Then, we argue that semantic compression provides a parsimonious unifying explanation of a wide range of experimentally observed memory errors. Finally, we present the results of a computational experiment which addresses biases of humans when they are tasked with reconstruction of earlier sketch drawings \cite{Carmichael1932}.

\section{Framework}

\subsubsection{Rate distortion theory}

The branch of information theory that deals with lossy compression is rate distortion theory (RDT). Optimal encoding in RDT is based on a knowledge of the statistics of the data. A central insight of RDT is that in the case of lossy compression there is no single optimal encoding: a trade-off emerges between the memory resources that are used for storing a given observation (rate) and the amount of distortion in the recalled memory. This precludes a single optimal solution but for any memory capacity constraint, a minimal expected distortion can be established. The RD curve, 
$$
R(D)=Inf_Q (R_Q), s.t. D_Q<D,
$$
which designates the minimal rate for any given distortion, defines the range of possible optimal encoding schemes for a given distribution over observations. An equivalent formulation can be obtained for convex RD functions by a constrained minimisation of the distortion with fixed rate, which can be formulated using the Lagrangian cost function
$$
L = \min\; D+\beta R.
$$
Any compression method can be associated with a point on the RD plane, with optimal algorithms lying on the curve. Assuming the curve is strictly convex, every point on it can be identified with a single value of $\beta$, which is the local slope of the curve. Thus, $\beta$ directly corresponds to a particular point on the rate-distortion trade-off continuum: for example a high value of $\beta$ is associated with strong compression, yielding a low rate but high distortion.

The distortion term, the cost associated with each possible alteration of the memory trace, is defined as the expected value of the distortion, $d$, between the original, $x$, and the reconstructed observation, $\hat{x}$, so that $D=\mathbb{E}_x[d(x,\hat{x})]$. An optimal lossy compression algorithm will selectively prioritize information such that alterations that are inconsequential according to this measure are discarded first. However, the distortion measure, $d(x,\hat{x})$, is left unspecified in RDT. 

The most widely used distortion function in the engineering domain is the squared distance of the original and reconstructed images, and has also been recruited in recent approaches to image compression with generative models \cite{Balle2016c}. While convenient, this is a poor measure of how the human memory weighs deviations from observations: it can be a good approximation to other forms of distortions at high rates (i.e. when distortions are very small), but at low rates (high distortion regime) it results in blurry images severely degrading the identification of image content. 

The information bottleneck (IB) method \cite{Tishby1999} extends RDT such that it guides the choice of the distortion function. The IB method introduces the idea of relevant quantization: they argue that distortion should be defined by identifying the quantities that we are interested in. The relevant information is then defined by the mutual information between the encoding ($Z$) and the relevant quantities ($Y$) so that the loss to be minimized becomes:
$$
L_{IB}=-I(Z,Y)+\beta I(X,Z)
$$
Minimising the IB loss function corresponds to a distortion measure that prioritizes information proportional to its predictiveness regarding the relevant quantities $Y$, which is formally equal to the distance between the distribution of the relevant quantity given the observed data and the distribution of the relevant quantity given the encoding: $d(x,\hat{x})=KL[p(y|x) || p(y|z)]$.  

The IB method provides an algorithm for optimising the loss function but it is not feasible to apply to high dimensional naturalistic data. However, it has recently been shown \citeA{Alemi2016} that in a variational approximation to the objective, the deep variational information bottleneck (DVIB),  an unsupervised version of the objective corresponds to the loss function of an approximate generative model called the $\beta$-VAE, thus establishing a link between generative models and rate distortion theory.

\subsubsection{Semantic compression}

Efficient compression is based on a knowledge of the statistics of the environment. We argue that semantic memory, viewed as a probabilistic generative model defined over sensory variables, represents the best estimate the brain has of such environmental statistics. Furthermore, it provides latent variables that are shaped by stimulus statistics, rewards, tasks, and predictive success. These variables include lower level acoustic or visual features such as phonemes, or objects as well as abstract concepts such as what constitutes a good chess move or melody. In addition, they contain the information relevant for predicting future observations. As a consequence, the latent variables summarize the relevant part of sensory information for the brain and we propose that this is precisely the information that should be prioritised when memory resources are constrained.

In our formalisation, observations correspond to snapshots of the states of sensory variables at a given time or time interval. Sensory experience is then interpreted through inference of latent variables, $z$, which are then encoded as the memory trace:
$$
\hat{z}(x_{obs})=O_z[p(z\;|\;x=x_{obs})],
$$
where $O_z$ stands for a point estimate of the posterior distribution. Such an estimate formally corresponds to an episodic memory trace. During recall, since semantic memory is assumed to be a generative model over observed variables, it can be used to recreate an experience based on the memory trace by conditioning on the stored values for latent variables. This results in a predictive distribution over observable variables, a point estimate of which can be regarded as the representation point for the particular value of the latent:
$$
\hat{x}(x_{obs})=O_{x}[p(x\;|\;z=\hat{z}(x_{obs}))].
$$

\subsubsection{Variational autoencoder}

Implementing semantic compression requires a probabilistic latent variable generative model of the domain, where inference can be performed over the latent variables. Inference, that is calculating the posterior distribution, is typically intractable for complex generative models capable of handling naturalistic data, thus necessitating the use of approximations. Variational Bayesian inference is a scalable, generally applicable approach to this problem in which the true posterior distribution is approximated by a distribution from a simpler distribution family. Once such a distribution family is chosen, the main goal is to minimise the Kullback-Leibler divergence between the true and approximate posteriors: $\textrm{argmin}_{\phi}\;\textrm{KL}(q(z|\phi)\;||\;p(z|x)).$
While this term cannot be computed directly, it can be shown that maximising the evidence lower bound (ELBO), 
$$
\mathcal{L}(\theta,\phi,x)=\mathbb{E}_{z\sim q_\phi(z|x)}(\log p_\theta(x|z))-\textrm{KL}(q_\phi(z|x)||p_\theta(z)),
$$
also minimises the KL divergence. The first term of this objective is often called the reconstruction term, alluding to the fact that it penalises inaccurate reconstruction of the observation. The second term is usually viewed as a regularisation term, as it penalises complex conditional posteriors. In a variational autoencoder (VAE) \cite{Kingma2014}, the approximating distribution is typically of a simple form, such as a Gaussian, which is parametrised via a neural network.  

A further benefit of the VAE is that it allows us to relate semantic compression to rate distortion theory through the correspondence with the unsupervised DVIB. In this correspondence, the reconstruction term is identified with the distortion $D$ and the capacity limiting regularisation term is identified with the rate $R$. The sole difference between the objectives is that in the DVIB case, the regularisation term of the ELBO is multiplied by the scalar $\beta$, which can be identified with the $\beta$ in the Lagrangian form of the RD objective. This variation has also been introduced with a generative modeling motivation in the form of the $\beta$-VAE \cite{Higgins2017}. It is remarkable that these separate lines of argument lead to the same objective, and we build on this connection to establish semantic compression in the normative framework for lossy compression.

\section{Distortions in human memory}

\subsubsection{Consequences of semantic compression}

Storing the states of latent variables of a semantic world model instead of the original observations implies specific patterns of memory errors. Here we briefly discuss these errors and contrast them with those described in the experimental literature on human memory and argue that semantic compression offers a parsimonious normative explanation for a large variety of memory errors and biases. Finally, we give a detailed demonstration of a specific effect observed by \citeA{Carmichael1932}. 

Since encoding of a memory trace corresponds to performing probabilistic inference over the latent variables, semantic compression implies that the uncertainty in inference affects the accuracy of the recalled memory trace. Classical memory experiments have shown that providing even a concise context which aids the interpretation of otherwise strongly ambiguous stimuli can greatly increase retention accuracy. For example in the study of \citeA{Bransford1972a}, prose passages are presented to subjects that are highly abstract descriptions of relatively simple situations such as washing clothes. The high level of abstraction renders the interpretation ambiguous therefore providing a simple context (such as a topic or a hand drawn image) before, but not after, encoding greatly increases recall score. Similar results for abstract nonsensical drawings were obtained \cite{Bower1975}, where a short interpretation substantially enhanced recall accuracy.

If the states of latent variables are stored instead of the sensory experience, then experiences attributed to the same latent state will be confused in memory. For latent variables such as sound textures, this corresponds to compressing sensory experience into sufficient statistics for the particular stimulus. This interpretation can be used to explain seemingly paradoxical results in auditory memory \cite{McDermott2013a}: in a task where audio excerpts have to be discriminated, the authors found that performance decreases with excerpt duration when the excerpts have the same sound texture, even though the key determinant of discrimination, information content, is strictly increases with excerpt length. The effect is reversed when the excerpts belong to different textures.

Due to the fact that the statistical model of a particular stimulus set provides the basis for optimal compression, the accuracy of this model affects the efficiency with which the relevant statistics can be extracted from observations. Expertise in a cognitive domain results in a better estimate of the observation statistics and in representations that are better suited to tasks in that domain. As a consequence,  recall performance is expected to depend on how well a particular stimulus conforms the environmental statistics. Indeed, \citeA{Baddeley1971} show that word recall performance increases as a function of the order to which the words conform to the statistics of the English language. Further, in a task where chess table positions have to be reconstructed by subjects, \citeA{Gobet1996} show that the number of pieces recalled correctly increases as a function of chess skill, however the difference between experts and non-experts is reduced for randomised board positions that do not conform to statistics observed during games.

Recall in semantic compression is a reconstructive process, where the generative model is conditioned on the stored memory trace. In case information about some features were lost during encoding, semantic memory can complement available information by relying on the prior distributions of these features. This results in a gist-like reconstruction of the stimuli, where values of not retained features are substituted with what is likely to have been part of the observation. Good examples of such false memory effects include the DRM effect \cite{Roediger1995}, where a long list of strongly related words has to be remembered. On a recognition test where 'lure' words are presented along with observed words, the authors find that when the lures are strongly related, they are falsely recognised as often as the originally presented words. Another example is the boundary extension effect  \cite{Intraub1989}: in a task where subjects have to redraw photographs from memory, they robustly recall surrounding regions not visible on the presented photograph, filling in unobserved but likely details in the scene.

\begin{figure}[htbp]
\begin{center}
\includegraphics[width=0.7\linewidth]{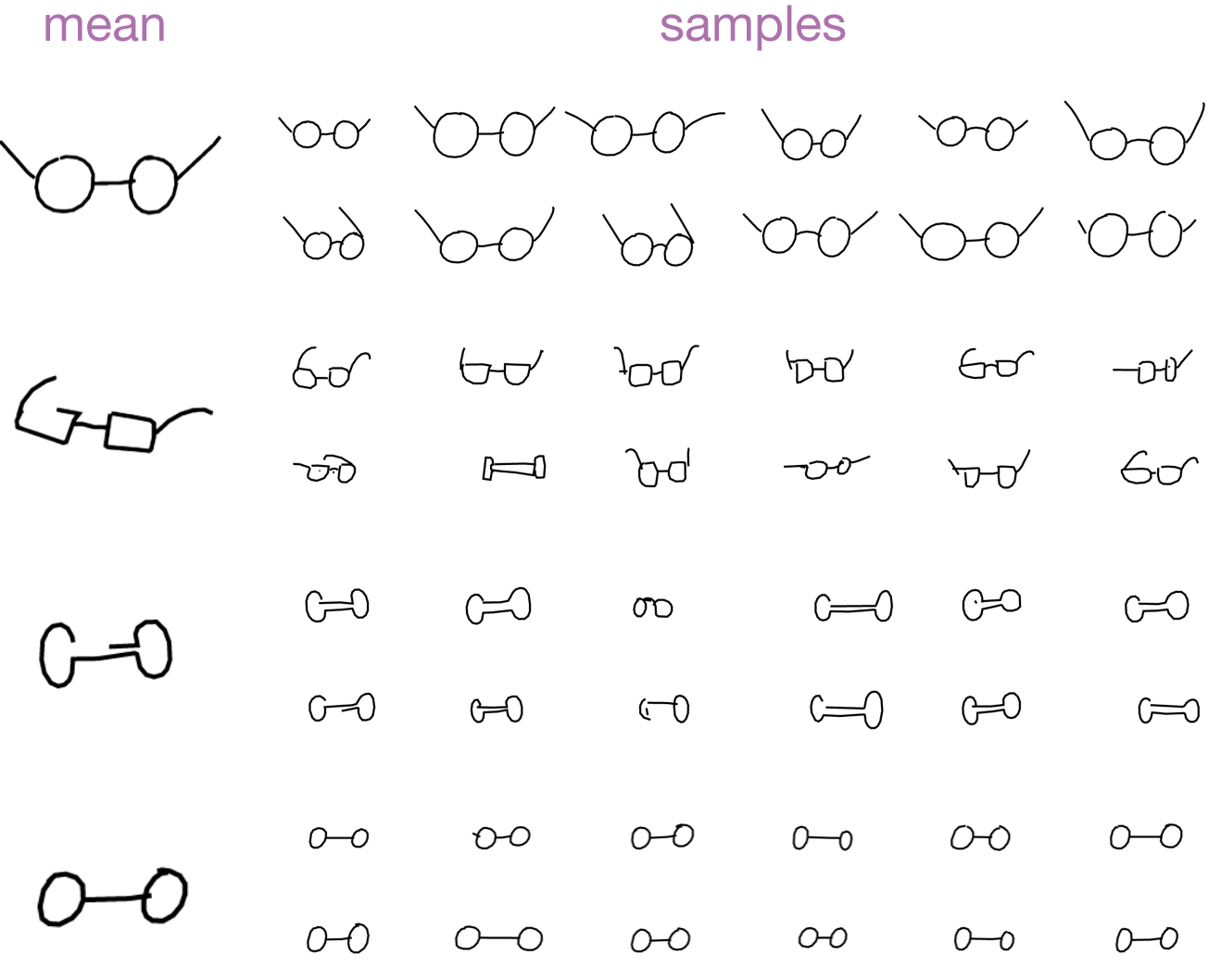}
\end{center}
\vspace{-0.4cm}
\caption{Learned representation. Some component means are presented along with samples. Intra-component differences are deemed smaller by the learned distortion function than inter-component differences, capturing human-like semantic distortions.
}
\label{fig:component_means}
\end{figure}

\begin{figure*}[!ht]
\begin{center}
\includegraphics[width=\textwidth]{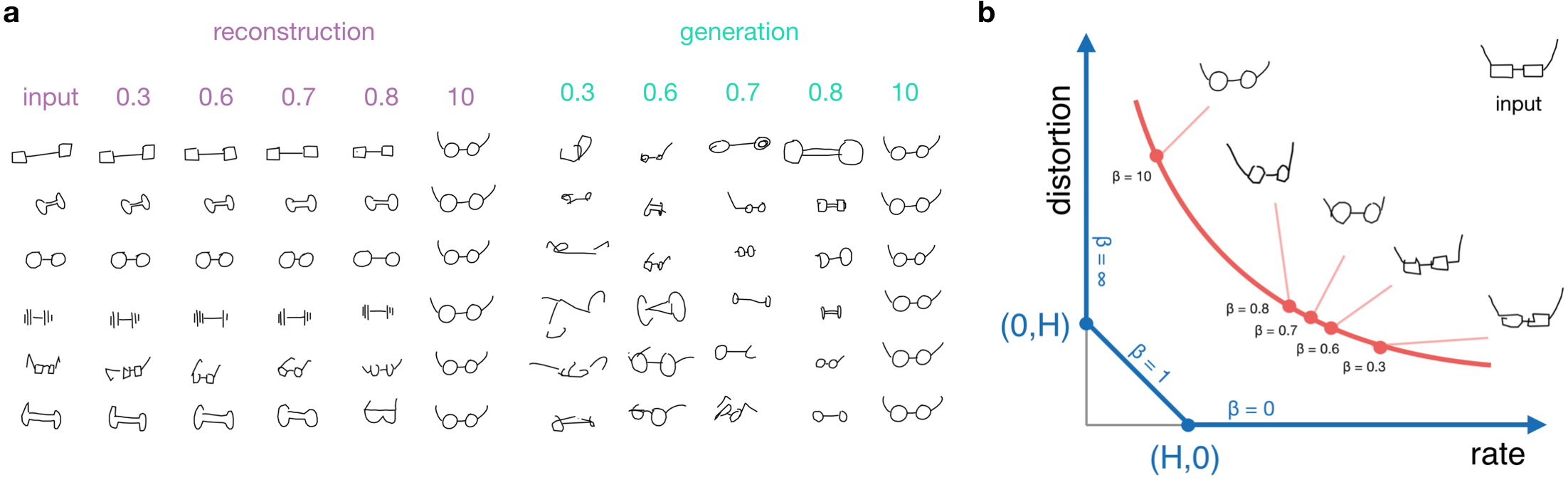}
\end{center}
\vspace{-0.4cm}
\caption{Reconstruction at different rates. a) Top row: value of $\beta$. Reconstruction: samples from the model with the given input in the left column. Generation: samples from the model without inputs. With increasing $\beta$, we lower the rate of compression: in reconstruction, idiosyncratic details of the input are lost. At the same time generation improves but also becomes less variable, at $\beta = 10$ producing one prototypical example. b) Blue curve: theoretical limit for no restrictions on parametric model family. Red curve: RD curve achievable by restricting posteriors to a parametric family such as in the sketch-rnn model. With increasing rate, compression is more faithful, while with decreasing rate, details are lost, rectangular shaped eyeglasses turn into more generic circular shaped ones.
}
\label{fig:beta_sweep}
\end{figure*}

\begin{figure*}[ht]
\begin{tabular}{c c}
\begin{minipage}{10.5cm}
\begin{center}
\includegraphics[width=10.8cm]{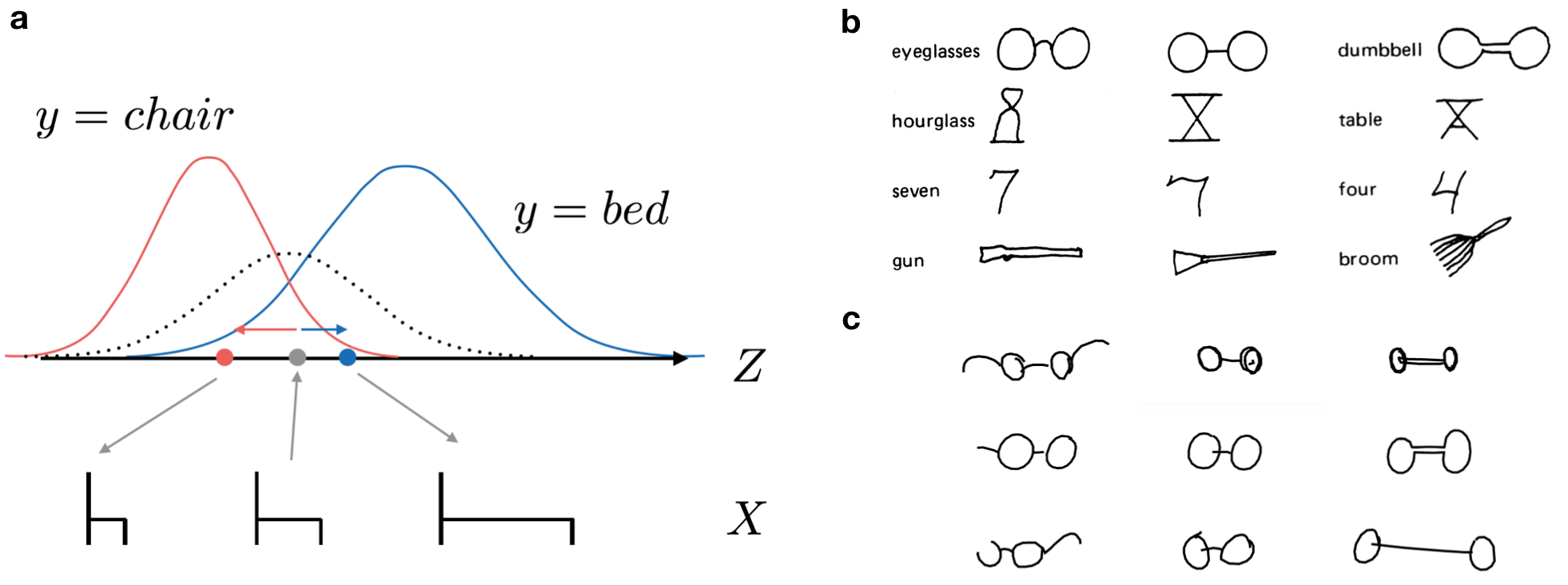}
\end{center}
\end{minipage} &
\begin{minipage}{6.7cm}
\caption{Memory distortions. a) Category information manifest in priors over the memory trace in the space of the latent $Z$ (red and blue lines). Combined with the likelihood term provided by the memory trace (grey dotted line) induces biases: recall without category label (grey dot) becomes distorted (red and blue dots) b) Examples from Carmichael et al. (1932). Middle column: figures shown to subjects. Left and right columns: distorted reconstructions by participants who received the corresponding category label. c) Our reconstruction of the memory distortion using the sketch-rnn model on eyeglasses and dumbbells. 
}
\label{fig:carmichael_effect}
\end{minipage}
\end{tabular}
\end{figure*}

\subsubsection{Memory for sketches}

In the following, we present a computational approximation to semantic compression in the domain of sketch-drawings under conditions where memory biases are known to emerge in human observers \cite{Carmichael1932}. We use this computational model to investigate whether it can enable efficient compression at multiple rates along the RD curve and demonstrate how it can explain systematic biases in human memory.  

In the classical experiment of \citeA{Carmichael1932}, intentionally ambiguous hand drawn sketches of objects from common categories were presented to subjects who were asked to reproduce these images after a given amount of delay. Two separate groups of participants received different category names along with the drawings. Depending on the categorical cue, systematic biases were introduced in reproduced images (Fig.\ 3b).

As an approximation of the semantic model for sketch drawings, we use the sketch-rnn architecture \cite{Ha2017} and the $\beta$-VAE objective. We trained this model on a dataset containing millions of sketch drawings of specific object categories that has recently become available in the Google QuickDraw dataset \cite{Ha2017}. To contrast distortions introduced by semantic compression with reproduction biases revealed by the Carmichael experiment, we selected potentially ambiguous object-pairs from the data set and trained the model on 75000 drawings of each category. Although the QuickDraw data set contains numerous object categories and rich naturalistic samples from every category, characteristics of recording the doodles preclude a large number of object pairs from the analysis. The QuickDraw data was recorded as part of a web browser game, where subjects had 20 seconds to draw an exemplar of a given category. However, if the drawing gets to a stage where an algorithm is able to recognise it as belonging to the provided category, a new trial is initiated. As a result, the data set contains a large number of half-finished drawings. Another limitation of the data set is that participants tend to draw prototypical exemplars of the category thus limiting the variance of the samples compared to natural hand drawings. This means that some of the designs appearing in the Carmichael experiment are not present in the data set and thus the model is oblivious to their interpretation. 

Since training of sketch-rnn is unsupervised, category labels can not be integrated during inference. We introduced these categories by fitting a mixture of Gaussians (MoG) model on top of the latent representation. The MoG model had 40 components with spherical covariances, where the number of components was chosen to allow separate components for varying drawing styles and drawing methods of the objects. While it might seem tempting to use a single component per category, the average of the variations of eyeglass drawings and those of a dumbbell are remarkably close to each other. The covariance of each component in the data is not hypothesised to be spherical, this constrained form was chosen solely to make the parameter estimation feasible. While the high dimensionality of the latent space precludes direct visualisation of learned distortions, concepts discovered by the MoG model can be investigated by generating drawings from each component (Fig.\ 1). Drawings that belong to the same component are considered close in the semantic representation, therefore the cost of confusing stimuli coming from the same component is relatively lower than that across components.

Fitting the model at different $\beta$ values, corresponding to different trade-offs between rate and distortion, results in qualitatively different behaviours (Fig.\ 2).  At high rates, compression behaves similarly to a completely episodic system: latents attempt to capture idiosyncratic details of the input but there is very limited generalisation and the semantic model learned in this regime is not capable of producing realistic unconditional samples. At lower rates, recall becomes similar to a completely semantic system: it leans increasingly on reconstruction via the predictive semantic model rather than retaining details of the observation. Note, that at an extremely low rate ($\beta=10$), latents become independent of the actual observation, generating a likely observation based on the marginal statistics of the data. This behaviour follows from the fact that maximum likelihood training and thus the ELBO objective does not give an explicit constraint on the latent representation (see \citeA{Alemi2017} for details). Actual reconstructions in the Carmichael experiment suggest that human semantic memory of sketches in this experimental setting is best represented at some intermediate rate. We have selected the $\beta=0.7$ model as our estimate: while capturing the level of variability in human memory experiments would require a larger $\beta$, increasing $\beta$ results in diluting category-related structure, suggesting that at high $\beta$ levels the sketch-rnn architecture is a poor approximation of human semantic representation. Changes in the level of compression are reflected in the increasing width of the posteriors at higher compression rates (data not shown). Hence, we emulated a higher compression level by inflating the posterior widths by a factor of 10, thus increasing compression level while leaving the representation of categories intact. 

To demonstrate that semantic compression results in the same kind of reproduction biases revealed by the Carmichael experiment, we trained the model on sketches of specific object pairs and selected potentially ambiguous sketches. When performing inference, we incorporate the category label provided in the experiment by conditioning on the sketch being generated from the category. According to the principles of Bayesian inference the category prior introduces a bias in the encoding (Fig.\ 3a), which will also be apparent in the generated drawing, qualitatively matching the bias introduced in the Carmichael experiment (Fig.\ 3c).

\section{Discussion}
In this paper we gave a normative argument for compressing events in human memory using the latent variables of semantic memory formalised as a probabilistic generative model of the environment. We argued that in the framework of information theory this corresponds to using the conditional likelihood of the model as the basis for the distortion function. This correspondence enabled us to integrate recent results in machine learning with memory research to make predictions on complex, naturalistic data. Our formalisation can parsimoniously explain a variety of memory biases, and here we gave a detailed demonstration of a classic example in the domain of reproduction of sketch drawings.

The relevance of RDT for explaining errors and biases in human visual working memory has recently been pointed out by \citeA{Sims}; the main difference between the approaches is that whereas they try to infer the distortion function in a bottom-up fashion from behavioural data on low dimensional perceptual tasks, we are attempting to give a normative argument for the appropriate form and comparing its predictions in high dimensional natural memory tasks. \citeA{Hemmer2009} use a dual-route generative model to explain the effect of semantic memory in a scene recall task, but they do not relate their method to compression. In our treatment dual routes are not required, as RDT provides a principled, continuous trade-off between episodic-like and semantic-like memory traces. \citeA{Gregor2016} explores using VAE-s trained with a generative objective for image compression in a machine learning context, and introduce an architecture for achieving different rates with the same model by conditioning on different levels of a hierarchy, however they do not relate their method to RDT or human memory.

\subsubsection{Acknowledgements}
The authors thank the anonymous reviewers for useful comments and Ferenc Husz\'ar for discussions. This work has been supported by the National Research,
Development and Innovation Fund of Hungary (Grant No. K125343) and an MTA Lend\"{u}let Fellowship. 

\bibliographystyle{apacite}
\setlength{\bibleftmargin}{.125in}
\setlength{\bibindent}{-\bibleftmargin}

\bibliography{mendeley_cleaned}

\end{document}